\documentclass[12pt,preprint]{aastex}

\begin{document}

\title{NEAR-INFRARED SPECTROSCOPY OF BROWN DWARFS: METHANE AND THE
TRANSITION BETWEEN THE L AND T SPECTRAL TYPES}

\author{{Ian S. McLean\altaffilmark{1},} \\ L. Prato\altaffilmark{1},
Sungsoo S. Kim\altaffilmark{1},
M. K. Wilcox\altaffilmark{1,2},
J. Davy Kirkpatrick\altaffilmark{3},
Adam Burgasser\altaffilmark{4}}

\altaffiltext{1}{Department of Physics and Astronomy,
University of California, Los Angeles, CA, 90095-1562}
\altaffiltext{2}{Department of Physics and Astronomy,
University of Hawaii, Hilo, HI, 96720}
\altaffiltext{3} {Infrared Processing \& Analysis Center, M/S 100-22, Caltech,
Pasadena, CA 91125}
\altaffiltext{4} {Division of Physics, Caltech, Pasadena, CA 91125}

\begin{abstract}

We report the possible identification of weak methane spectral features 
in the near-infrared $K$ band in two late L dwarfs, DENIS 0205-11 (L7)
and 2MASS 1523+30 (L8). New, high signal-to-noise ratio flux-calibrated 
spectra, spanning the wavelength interval
1.10$-$2.35 $\mu$m~ with an average resolving power $R = 1800$ were obtained 
using NIRSPEC on the Keck II telescope. Results are reported and compared for
three late L dwarfs (L5, L7, and L8) and two T dwarfs (T1 and T6). 
The spectra, which are continuous through the
atmospheric absorption bands, show the development of deep steam
bands and the weakening of iron hydride features
through the L dwarfs and the emergence of strong
methane bands in the T dwarfs. A detailed comparison of the
$K$ band regions with synthetic spectra suggests that the
weak features seen in the L7 and L8 dwarfs at
2.20 $\mu$m are likely to be methane.  We see no evidence for
methane in the $H$ band.
At the $R = 1800$ resolution, significant differences are evident
between the spectral signatures of the L8 and the
T1, leaving room for additional transition objects (L9 or T0). 

\end{abstract}


\section{Introduction}

Very low-mass stars and brown dwarfs with effective temperatures below
$\sim$2200 K exhibit complex spectra rich in molecular features.
Many such objects have now been identified, and the L and T classes, the first
significant, new spectral classes in almost a century, have been added to the
familiar OBAFGKM sequence \citep{kir99, mrt99, brg01}.
Spectroscopically, the characteristics of L and T dwarfs are significantly
different from late M dwarfs \citep{geb96, rui97, tin97, kir99, rei01,
brg01}.  Titanium oxide and vanadium oxide bands lose their prominence in
L dwarfs, revealing absorption by iron hydride (FeH) and chromium hydride.
The alkali metals (Na, K, Rb, and Cs) become very strong and the 
pressure-broadened wings of these lines
are a major opacity source at visible wavelengths \citep{brw00}.
In the infrared (IR), strong steam (H$_{2}$O) bands appear and become
deeper in lower temperature objects. These IR H$_{2}$O bands can
be used for spectral classification even at resolving powers
of $R=\lambda/\Delta\lambda \sim 100-800$
\citep{mcl00b, rei01, tes01, brg01},  which is advantageous since
brown dwarfs are much brighter at near-IR wavelengths.
For example, for the L4 dwarf GD 165B, I $=$ 19.6 mag
but $K\sim$14 mag \citep{zuc92, kir99}.
Classification as a T dwarf requires that
strong methane (CH$_4$) absorption be present in
the $J$, $H$ and $K$ bands \citep{kir99},
although \citet{nol00} have shown that weak
CH$_4$ appears in objects as early as L5 dwarfs when observed 
at the fundamental band at $\sim$3.3 $\mu$m.

With the recent availability of powerful, new IR spectrometers on 
very large (8$-$10 m) telescopes, higher
resolution spectroscopy of low-mass stars and
brown dwarfs is now possible. In this paper we present new results from
NIRSPEC, the first high-resolution cryogenic 1$-$5 $\mu$m~
spectrometer for the Keck II 10-meter telescope. This work is part of a larger
survey \citep{mcl01} called the NIRSPEC Brown Dwarf Spectroscopic Survey
(BDSS). The goals of the survey are
({\it i}) to obtain a consistent set of IR spectra with $R \sim 2000$
for a large sample of very low-mass stars and brown dwarfs for
comparison with spectral energy distribution
models, ({\it ii}) to obtain a set of
very high-resolution spectra ($R\sim 20,000$) for detailed comparison of
individual spectral features with model
atmosphere predictions, and ({\it iii}) 
to monitor selected sources for Doppler shifts
induced by unresolved binary companions.

In this paper, we present high signal-to-noise, flux-calibrated
spectra covering 1.10$-$2.35 $\mu$m~ 
at a resolution of $\sim$ 1800 for five objects spanning the 
transition between the L and T spectral types. We report the probable 
discovery of CH$_4$ features in the $K$ band of L7 and L8 dwarfs.
The observations, data reduction and calibration methods are described in 
\S 2. In \S 3 we present the new results and discuss their implications, with 
particular emphasis on the transition between the L and T classes.
Section 4 is a summary of our conclusions.

\section{Observations and Data Reduction}

Table 1 lists the objects discussed in this paper and summarizes their
photometric properties. Preliminary results for four of the objects
were reported previously by \citet{wil00}, but the
data have been re-analyzed and flux-calibrated. The fifth object is an 
early T dwarf discovered by the Sloan Digital Sky Survey \citep{leg00}. 
The L5 and L7 are non-lithium objects whereas the L8 
is a lithium brown dwarf \citep{kir99, kir00}. Four targets
were observed in 1999 June$-$August, and one in 2000 December.
The spectral types listed in Table 1 are from \citet{kir99} and
\citet{brg01}.

The Keck II Near-IR Spectrometer, NIRSPEC, is a cross-dispersed, 
cryogenic echelle spectrometer employing a 1024 $\times$ 1024 ALADDIN 
InSb array detector \citep{mcl98, mcl00a}. In echelle mode the 
resolving power is $\sim 25,000$~ (2 pixels). A low-resolution mode of
$R\sim 2000$~ (2 pixels), can be obtained by replacing the echelle 
grating with a plane mirror. For the present study, we used
the lower resolution mode to observe a broad wavelength span in the
minimum number of grating settings and the
shortest time. The total wavelength region
covered was 1.10$-$2.35$\mu$m~, with a
spectral resolving power of $R = 1800$ (6$-$13 \AA~ resolution),
corresponding to a slit width of 0.$''$38 (2 pixels).  Seeing
conditions generally gave full-width half 
maximum values of 0.$''$5 or better. The entire
set of observations represents about 6 
hours of on-source integration time. 
For each object, integrations of 600 s were taken at
two nod positions, separated by $\sim$20$''$ along the 42$''$ slit.
NIRSPEC's IR slit-viewing camera
was used frequently during the spectroscopic exposures to check centering
on the slit while offset-guiding on a field star. Stars of spectral type 
A0 V $-$ A2 V were observed at the same
airmasses as the target objects to calibrate
for absorption in the Earth's atmosphere. Neon and argon arc lamp spectra,
together with the spectrum of a flat field lamp and dark frames, were taken
immediately after the observation of each source.

Since the NIRSPEC spectra are slightly curved and distorted
by the high-throughput optics, it is necessary to correct for
this effect before applying standard extraction techniques.
We (S. S. Kim, L. Prato, and I. McLean) have developed
software at UCLA to rectify the data and maximize
the efficiency of the processing overall. This code
utilizes the spectral traces of the bright early-type standard star and known 
arc lamp lines to map the distorted image to a spatially and spectrally
linear, rectified image.
First, the standard star (used to calibrate 
telluric atmospheric absorption), the target
object, and the flat field (created by
subtracting off the dark frame from the
flat field lamp frame) images are cleaned, replacing
bad pixels by interpolating over their neighbors. After rectification, the
pairs of nodded target frames are subtracted to remove background and
divided by the flat field.  Comparison of the target spectrum with the
OH night sky spectrum is used to detect artifacts caused by incomplete
subtraction and saturated OH lines.  At this point, the
power spectrum of each source is examined for evidence of
fringing effects within the instrument; these
effects are generally weak in the low-resolution mode.
Spectra were extracted by summing $\sim$10
adjacent rows, assuming that the noise is dominated by Poisson statistics.
The same procedure is applied to the telluric standard star,
except in this case any intrinsic stellar
features, such as the Brackett and Paschen lines,
are carefully removed from the spectra
by interpolation. We then divide the target spectra
by the corrected telluric standard spectra
and multiply by a blackbody equivalent in temperature to that of the
standard star \citep{cox00}
to maintain the true form of the target spectra. Finally, the 
spectra produced from the two nod positions are averaged.

To flux-calibrate the data, we use 2MASS photometry except for the Sloan 
object, in which case the photometry is
from \citet{leg00} on the UKIRT system.
Band-averaged flux densities for Vega are defined by
\begin{equation}
\label{flux_vega}
F^{band}_\lambda ({\rm Vega}) \equiv { \int F(\lambda ; {\rm Vega})
S(\lambda) \, d\lambda \over
\int S(\lambda) \, d\lambda} \, ,
\end{equation}
where $F(\lambda ; {\rm Vega})$ is the flux density of Vega and
$S(\lambda)$ is the transmission profile of a certain bandpass which
is the product of detector quantum efficiency, filter transmission,
optical efficiency, and atmospheric transmission. Band-averaged flux
densities for targets are similarly defined:
\begin{equation}
\label{flux_target}
F^{band}_\lambda ({\rm target}) \equiv { \int F(\lambda ; {\rm target})
S(\lambda) \, d\lambda \over
\int S(\lambda) \, d\lambda} \, .
\end{equation}
After removing the intrinsic telluric features, the flux density of the
target is related to the data number count (after the whole reduction above),
$D(\lambda)$, by a proportional coefficient $c$ such that
\begin{equation}
\label{F2D}
F(\lambda ; {\rm target}) = c \, D(\lambda) \, .
\end{equation}
Using Vega to define zero magnitude at all wavelengths, the
magnitude of the target at a given band becomes
\begin{equation}
\label{mag}
{\rm mag}({\rm target}) =
-2.5 \log {F^{band}_\lambda ({\rm target}) \over
F^{band}_\lambda ({\rm Vega})} \, .
\end{equation}
Then one can flux-calibrate a spectrum by finding $c$ from
\begin{equation}
\label{c}
c = {\int S(\lambda) \, d\lambda \over \int D(\lambda) S(\lambda)
\, d\lambda} \, F^{band}_\lambda ({\rm Vega}) \, 10^{-0.4 \,
{\rm mag} ({\rm target})} \, .
\end{equation}
When using the 2MASS $J$, $H$,
and $K_s$ magnitudes for the flux-calibration,
$S(\lambda)$ in the above equation is represented by the
transmission profile of the 2MASS bandpasses, which we obtained from
{\sf
\mbox{http\,://www.ipac.caltech.edu/}\linebreak[0]\mbox{2mass/}
\linebreak[0]\mbox{releases/}\linebreak[0]\mbox{second/}
\linebreak[0]\mbox{doc/}\linebreak[0]\mbox{sec3\_1b1.html}}.
Our $F^{band}_\lambda ({\rm Vega})$ values are based on data
from \citet{ber95}, kindly provided by D. Saumon, and
the 2MASS filter set. We also compared these
values to fluxes from \citet{coh92},
which are based on the UKIRT filter set and the
atmospheric absorption at Kitt Peak \citep{cox00}. A 5\% difference
was found in the $J$ band, but only 1-2\% difference in the
$H$ and $K_s$ bands.
For the flux calibration of the T1 dwarf, we used the
IRCAM/UFTI $J$, $H$, and $K$ filter bandpasses, available on
the UKIRT website, with the $F^{band}_\lambda ({\rm Vega})$ values
from \citet{coh92}.
The resulting spectra are shown in Figure 1. A gaussian smoothing function has
been applied in this representation.
Based on counting statistics, the signal-to-noise ratio is 
typically $>$100. Except in the center 
of the deep H$_2$O bands where atmospheric
transmission is poor, most of the
small-scale structure in the spectra is real.

\section{Results and Discussion}

From Figure 1, it is evident that the absolute flux in the $J$ band
decreases relative to that in the $H$ band
from L5 $-$ L8, presumably influenced by changes in dust opacity 
\citep{all01, tsu01, mrl01, bur01},
but recovers dramatically
in the T dwarfs as the dust settles out below the photosphere.
This effect is apparent in the $J-H$ colors of the objects;
the L8 is clearly the reddest, with a $J-H$ of 1.32 (Table 1).
H$_2$O bands deepen from L5$-$T5; the depth
of these bands is a useful indicator of spectral class
\citep{mcl00b, rei01, tes01}.
A strong FeH band at 1.19 $\mu$m weakens
from L5 $-$ L7 and is gone by L8. The equivalent
widths of the neutral potassium lines (K I)
change slightly as the lines at first weaken through L8 and
then appear to increase in strength at T1 before declining again. 
At least part of the variation in the K I lines is a result of
pressure-broadening and the apparent strengthening in lower
temperature objects is likely caused by
the different depth in the atmosphere that
is reached when dust is absent \citep{sau00}.
At 2.295 $\mu$m, the CO v$=$2$-$0 bandhead is prominent
through L8, but other CO transitions in this region are not. 
The presence of CH$_4$ in the early T dwarf \citep{leg00}
produces distinct features in
the $J$, $H$ and $K$ band spectra.
In cooler objects such as the T6, these features develop into large
absorption troughs easily detected in low resolution spectra.
No evidence of the $4s^{2}S-4p^{2}P^{0}$ doublet of Na I 
at 2.2076 $\mu$m~ is apparent, even in our 
L5 spectrum.  Although the opacity of this doublet is still
strong for objects with T$_{eff} = 1600$ K, it is insignificant
in comparison to the continuum absorption by H$_2$O and H$_2$,
consistent with our non-detection (D. Saumon 2001, private
communication).

Given the appearance of CH$_4$ in the late L dwarfs
at 3.3 $\mu$m \citep{nol00, geb01}, we searched
for CH$_4$ at shorter
wavelengths. At the smoothed resolution of
Figure 1 (R$\sim$900), there are features 
in the L7 and L8 dwarfs at 1.67 $\mu$m~ and 
2.20 $\mu$m, close to where CH$_4$ is expected to appear.
The distinct $H$ band line, prominent in Figure 2
at $\sim$ 1.669 $\mu$m, is not the 2$\nu_3$ CH$_4$ feature
\citep{leg00} but rather an artifact of incomplete removal of a
saturated OH night sky line \citep{cox00}.  These telluric lines
may also be the source of similar features seen in the L dwarf
spectra of \citet{rei01} and \citet{brg01}.
Our observations are compared to models 
with and without CH$_4$ opacity, kindly supplied by D. Saumon.
The models shown in Figures 2 and 3 
have T$_{eff} = 1500$ K with 
$log(g) = 5$, solar metallicity, and $f_{rain} = 5$ (D. Saumon and
M. Marley 2001, private communication; Ackerman \& Marley 2001).
CH$_4$ is present 
in the model near 1.667 $\mu$m but no corresponding feature
is evident in the L dwarfs.  The T1 dwarf does present a small
absorption feature at 1.667 $\mu$m and a shallow, broad dip in
the continuum from $\sim$ 1.665 $-$ 1.675 $\mu$m (Figure 2).
No differences were apparent, at a level greater than the 
average noise in our data, in
the CH$_4$ and non-CH$_4$ models
around the 1.32 $\mu$m region of the $J$ band.

Figure 3 shows the expanded $K$ band region. 
From 2.05$-$2.15 $\mu$m the dominant
features are unresolved transitions of
H$_2$O. There is excellent correspondence
among the three L dwarfs in this region; most of the strongest
features are also present in the T dwarfs, the coolest of which are
well fit by models
(Saumon et al. 2000; D. Saumon 2001, private communication). 
At the other end of the
$K$ band spectrum, the dominant feature is the CO
band head at 2.295 $\mu$m. On the long wavelength
side of the band head, the flux increases slightly but
fails to recover to the continuum level
in the L5 dwarf. The later L types show only a
decrease in flux at wavelengths longer than the CO bandhead, and for 
the T1, the CO absorption is blended with
weak CH$_4$. In general, for L dwarfs later than L5, there
is a gradual decrease in flux beyond 2.15 $\mu$m. This decrease has been
attributed to collisionally induced absorption
by H$_2$~ \citep{rui97, tok99}.
When CH$_4$ absorption occurs, there is a
break in the slope near 2.18 $\mu$m and a minimum
develops at 2.2 $\mu$m. This region is
shown by the shaded area in Figure 3. 
The L7 and L8 spectra display a
small break in the overall shape around 2.18 $\mu$m.  Sharp features 
appear near 2.20 $\mu$m, analogous to the lines in the
CH$_4$ opacity model.  Even in the L5 dwarf spectrum a small feature
at 2.2 $\mu$m suggests the presence of CH$_4$.
Hence, it appears that weak CH$_4$ absorption
is present in the latest L dwarfs.

Clear differences between the L8 and T1 spectra at this 
resolution are evident in Figure 3.
The L8 spectrum is more similar, both in overall shape
and in individual features, to the L7.
We therefore propose that identification of an additional
subclass object, such as an L9 or a T0, would provide a smoother 
transition between the L and T dwarfs.

\section{Conclusions}

We have presented a consistent set
of near-IR spectra, with a resolving power of
$R = 1800$, for a sample of five objects
cooler than $T_{eff} \sim 1600$. These data are part of our
larger on-going spectroscopic survey. High-quality, flux-calibrated spectra
from 1.13$-$2.33 $\mu$m are reported for three L dwarfs and two T dwarfs.
We have used the higher spectral resolution
to search for early evidence of the 
onset of CH$_4$ absorption in the late L dwarfs.
By comparing the observations to spectra from synthetic model atmospheres with 
and without CH$_4$ opacity, we associate
the 2.2 $\mu$m features
with CH$_4$ indicating that methane
first appears in dwarfs as early as L7 in the $K$ band.
The observations presented here show that the spectral
distinction between L8 and T1 is significant at this resolution, 
therefore leaving room for additional transition objects.

\acknowledgments

It is a pleasure to acknowledge the hard work of past and
present members of the NIRSPEC instrument team, without whose
efforts these observations would not have been possible.
It is also a pleasure to acknowledge CARA instrument specialists Tom
Bida and David Sprayberry, in addition to the CARA staff and Keck II
Observing Assistants for their exceptional support.  We thank
an anonymous referee for useful comments and suggestions which
improved the manuscript.
We are especially grateful to D. Saumon and M. Marley 
for extensive help and discussions about models, and for providing synthetic 
spectra.
Data presented herein were obtained at the W.M. Keck 
Observatory, which is operated as a scientific partnership
between the California 
Institute of Technology, the University of California and NASA.
The Observatory was made possible by the generous financial support of the 
W.M. Keck Foundation.


\clearpage

\begin{figure}
\plotone{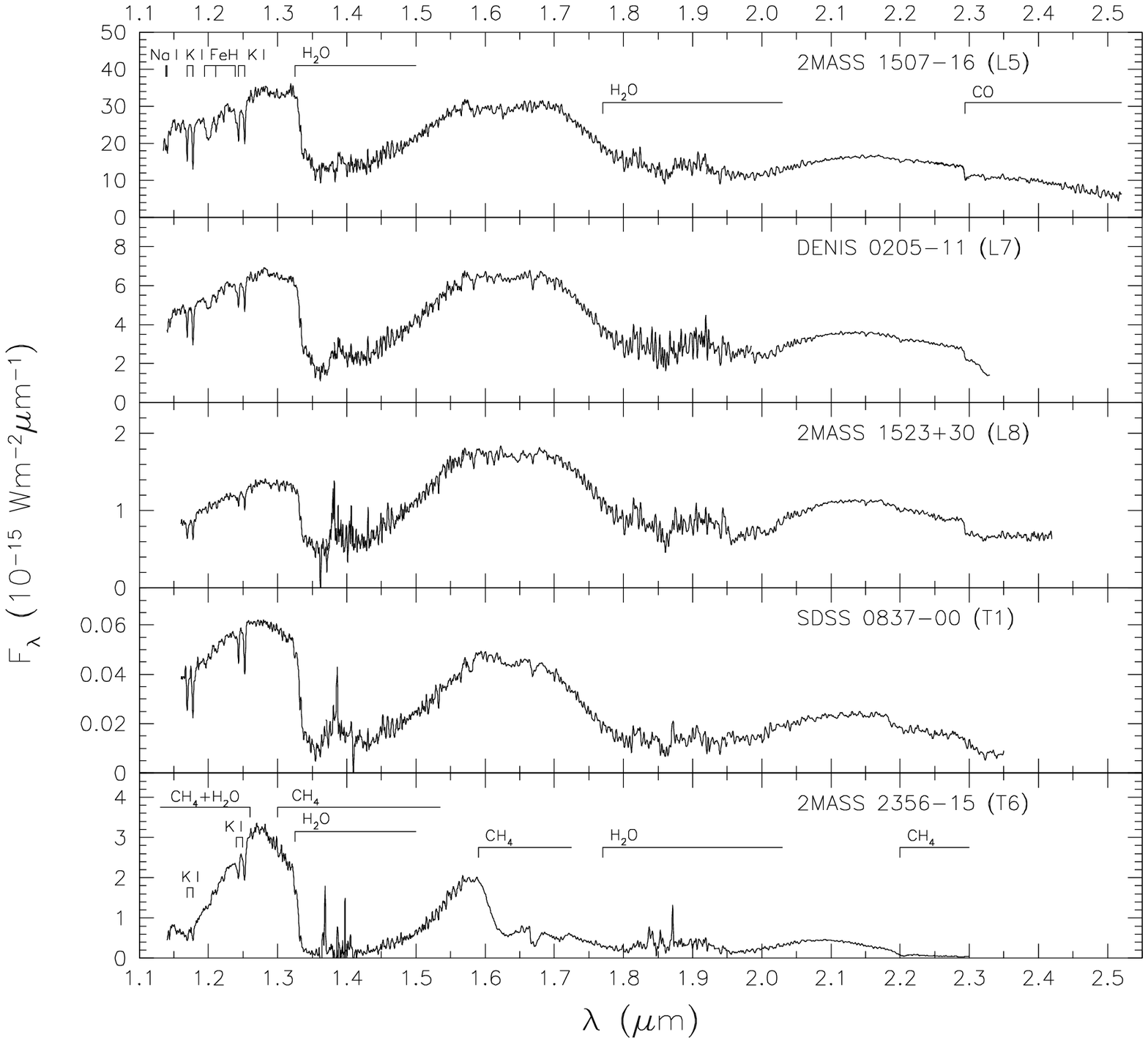}
\caption{Flux calibrated near-infrared
spectra of three L dwarfs and two T dwarfs obtained with NIRSPEC
on the Keck II telescope.  The original resolution ($R = 1800$) has 
been smoothed by a gaussian function of width 2 pixels, resulting
in $R = 900$ in these plots.  Some of the small, sharp features
in the $H$ band are attributable to saturated
night sky OH lines (see text).
\label{fig1}}
\end{figure}

\begin{figure}
\plotone{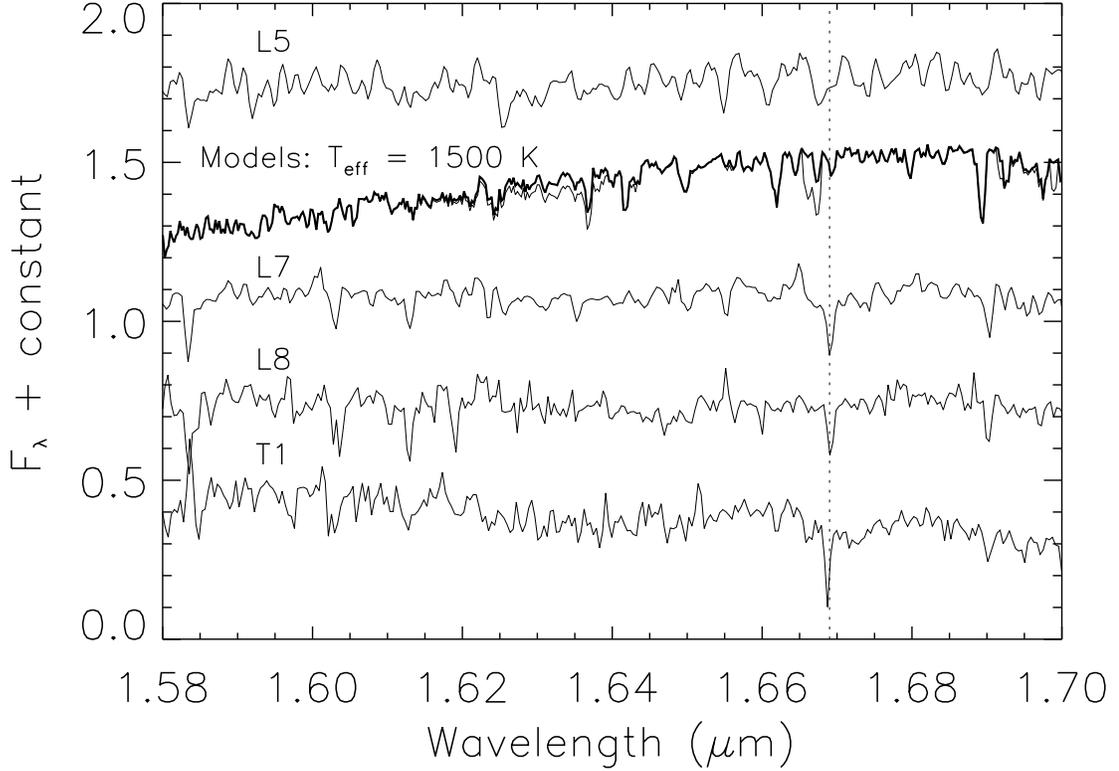}
\caption{An expanded view
of the $H$ band region for the first four objects in Figure 1.
The data are unsmoothed and $R = 1800$.
Each spectrum has been normalized and offset by a constant. The
dotted line indicates a spectral line associated with night sky OH near
the 2$\nu_3$ CH$_4$ feature. 
Also shown at similar resolution are two models (see text) 
courtesy of D. Saumon and M. Marley (2001, private communication),
one with CH$_4$ opacity and one without (boldface).  
In the L dwarfs, we see no
clear evidence of the 1.667 $\mu$m CH$_4$ feature present in the
model.  Although the OH line contamination is also present in the
T1 dwarf, a small CH$_4$ line appears to be developing at 
1.667 $\mu$m and a wider depression dips the continuum between
1.665 and 1.675 $\mu$m.
\label{fig2}}
\end{figure}

\begin{figure}
\plotone{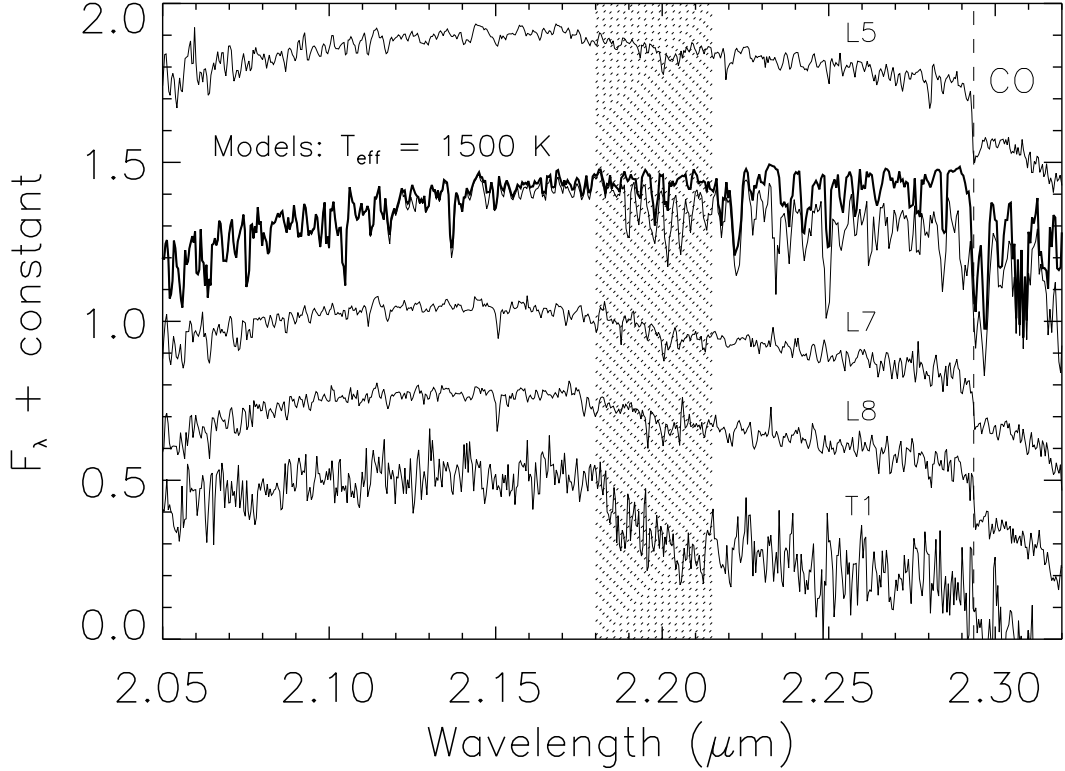}
\caption{Similar to Figure 2 but for the $K$ band region.
The data are unsmoothed and $R = 1800$.
Each spectrum has been normalized and offset by a constant. The
shaded region indicates where CH$_4$
absorption first appears in the T dwarf. Also shown at similar resolution
are two models (see text) with and without (boldface) CH$_4$
opacity.  Incompletely removed night sky lines are again apparent
at 2.195 and 2.150 $\mu$m.  However, the most
prominent CH$_4$ feature in the model, at 2.201 $\mu$m, a region 
unaffected by strong OH \citep{cox00}, is clearly present in the
L7 and L8 dwarfs.  The continuum longward of 2.180 $\mu$m is
strongly suppressed by CH$_4$ in the T1 dwarf.  A small feature
is present in the L5 dwarf at 2.2 $\mu$m; possibly revealing the initial
onset of CH$_4$ in brown dwarfs in the $K$ band.  \label{fig3}}
\end{figure}


\clearpage

\pagestyle{empty}

\begin{deluxetable}{lccccc}
\tablewidth{0pt}
\tablecaption{Photometric Properties of the Observed Objects \label{tbl-1}}
\tablehead{
\colhead{Name} & \colhead{Spectral Type} & \colhead{J (mag)} & \colhead{H (mag)}
 & \colhead{K$_s$ (mag)} & \colhead{Reference}}
\startdata
2MASSI J1507476-162738 & L5 & 12.82 & 11.89 & 11.30 & 1\\
DENIS-P J0205.4-1159 & L7 & 14.55 & 13.59 & 12.99 & 2\\
2MASSW J1523226+301456\tablenotemark{a} & L8 & 16.32 & 15.00 & 14.24 & 1 \\
SDSS J083717.21-000018.0\tablenotemark{b} & T1 & 16.90 & 16.21 & 15.98 & 3 \\
2MASSI J2356547-155310 & T6 & 15.80 & 15.64 & 15.83 & 4\\
\enddata

\tablenotetext{a}{Also known as GL 584C}
\tablenotetext{b}{Photometry on UKIRT system}

\tablecomments{References: (1) \citet{kir00},
(2) \citet{kir99}, (3) \citet{leg00}, (4) \citet{brg01}}

\end{deluxetable}

\end{document}